\RequirePackage{fix-cm}
\documentclass{svjour3}                     

\usepackage{color}
\usepackage{graphicx}
\usepackage{subfigure}
\usepackage[belowskip=-5pt,aboveskip=0pt,skip=6pt,labelfont=normalfont,textfont=normalfont]{caption}

\usepackage[british]{babel}
\usepackage[utf8]{inputenc}
\usepackage{latexsym}
\usepackage[T1]{fontenc}
\usepackage{mathptmx}
\usepackage{xspace} 
\usepackage{amsmath}
\usepackage{mathtools}
\usepackage{gensymb} 
\usepackage{enumerate}  

\usepackage{cite}
\usepackage[sort&compress,square,numbers]{natbib} 
\usepackage{hyperref}
\usepackage{pbox}
\definecolor{pdflink}{rgb}{0.0, 0.0, 1.0}
\definecolor{pdfcite}{rgb}{0.0, 1.0, 0.0}
\definecolor{pdfwww}{rgb}{1.0, 0.0, 0.0}
\usepackage{xspace} 
\DeclareMathOperator*{\argmin}{argmin}

\usepackage[nolist]{acronym}
\newcommand{\uCT}{$\mu$CT\xspace}
\newcommand{\Np}{$N_p$\xspace}  
\newcommand{\Nt}{$N_\theta$\xspace}  
\newcommand{\Nr}{$N_r$\xspace}  
\newcommand{\Nhtr}[0]{N_h,N_\theta,N_r}  
\newcommand{\MTFR}{$m_r$\xspace}  
\newcommand{\MTFT}{$m_\theta$\xspace}  
 \journalname{Journal of Nondestructive Evaluation}

\begin{document}
	
\DeclareGraphicsExtensions{.pdf,.png,.jpg,.svg}
\begin{acronym}
\acro{MR}{Magnetic Resonance}
\acro{CT}{Computed Tomography}
\acro{uCT}[\uCT]{Computed micro-Tomography}
\acro{UGCT}{Ghent University Centre for X-Ray Tomography}
\acro{RP}{Radiation Physics}

\acro{GPU}{Graphics Processing Unit}
\acro{SART}{Simultaneous Algebraic Reconstruction Technique} 
\acro{DART}{Discrete Algebraic Reconstruction Technique} 
\acro{FBP}{Filtered Back-Projection}
\acro{NN-hFBP}{Neural Network Hilbert transform based \ac{FBP}}
\acro{pixel}{picture element}
\acro{voxel}{volume element}
\acro{SOD}{Source Object Distance}
\acro{SDD}{Source Detector Distance}
\acro{ROI}{Region Of Interest}
\acro{Np}[\Np]{number of projections}
\acro{Nt}[\Nt]{number of azimuthal voxels}
\acro{Nr}[\Nr]{number of radial voxels}

\acro{MTF}{Modulation Transfer Function}
\acro{MTFR}[\MTFR]{radial MTF}
\acro{MTFT}[\MTFT]{azimuthal MTF}

\acro{NDT}{Non-Destructive Testing}
\acro{RHT}{Randomized Hough Transform}
\end{acronym}

\title{Fast tomographic inspection of cylindrical objects
}


\author{Wannes~Goethals \and
        Marjolein~Heyndrickx \and
        Matthieu~Boone
}


\institute{W. Goethals \at
              UGCT-Department of Physics and Astronomy, Proeftuinstraat 86, 9000 Ghent, Belgium \\
              \email{wannes.goethals@ugent.be}
           \and
           M. Heyndrickx \at
           \email{marjolein.heyndrickx@ugent.be}
           \and
           M. Boone \at
           \email{matthieu.boone@ugent.be}
}

\date{Received: date / Accepted: date}

\maketitle

\begin{abstract}
	
This paper presents a method for improved analysis of objects with an axial symmetry using X-ray \ac{CT}.
Cylindrical coordinates about an axis fixed to the object form the most natural base to check certain characteristics of objects that contain such symmetry, as often occurs with industrial parts.
The sampling grid corresponds with the object, allowing for down-sampling hence reducing the reconstruction time. This is necessary for in-line applications and fast quality inspection. 
With algebraic reconstruction it permits the use of a pre-computed initial volume perfectly suited to fit a series of scans where same-type objects can have different positions and orientations, as often encountered in an industrial setting. 
Weighted back-projection can also be included when some regions are more likely subject to change, to improve stability.
Building on a Cartesian grid reconstruction code, the feasibility of reusing the existing ray-tracers is checked against other researches in the same field.\\

\end{abstract}

\section{Introduction} 

Non-destructive evaluation, either in-line or on random test samples, is paramount in a modern production environment to detect internal faults at an early stage. For assembled products, a common fault case is the situation where a component is missing in the final product. These components are often concealed in the assembly, hindering visual inspection. X-ray based imaging may offer better insight on manufacturing quality. 

For in-line inspection, X-ray transmission scanning is usually applied to make 2D projection images. Common examples can be found in defect detection~\cite{haff2008x,zou2015automatic,xu2010automatic} or material identification~\cite{duvillier2018inline}. On the other hand, the more intensive 3D imaging by X-ray~\ac{CT} is mainly used off-line.
The first industrial applications of \ac{CT} can be found since the 1980s in the field of material analysis, e.g. observing inner structure and processes or locating internal defects.
More recently, \ac{CT} scanners are being used dedicated to metrology for dimensional quality control. High importance is given in this case to accuracy, traceability to the unit of length and measurement uncertainty~\cite{kruth2011computed,ametova2018computationally,jones2018limited}.
New developments in CT technology, driven by research at synchrotron facilities and medical imaging, allow the technology to be applied in an automated production environment too~\cite{shefer2013state,Thomas2016line}. \\

As the industry is gradually becoming aware of the options that X-ray CT offers as a non-destructive 3D inspection tool, this leads to the demand for systems specifically designed for scanning sequences of same-type objects in a high throughput environment. \\
The challenges lie in the fact that a variety of factors restrict the number of projections to \textit{as low as technically feasible}.
The solutions can be found in more advanced reconstruction techniques like discrete reconstruction~\cite{batenburg2011dart} or neural-network based solutions~\cite{janssens2018neural}.
Even though iterative reconstructions can't compete with the analytic counterparts in terms of reconstruction time for standard cone-beam geometry, they allow making fewer projections with still reasonable results by incorporating prior knowledge. This reduces the scan time and the absorbed radiation dose per object. \\

Previous studies have shown that iterative reconstruction schemes using cylindrical coordinates can alleviate memory requirements and be less subject to noise than \ac{FBP} algorithms~\cite{Thibaudeau}. As a negative point of attention, Jian et al. \cite{Jian} reported a loss in resolution due to non-uniform sampling. 
In these studies, the cylindrical coordinate axis is positioned along the vertical axis of the rotation stage.
\\

In this work the specific choice is made not to fix the reconstruction axis to the vertical rotation axis, but to attach the coordinate system to the detected symmetry axis of the scanned object.
In an industrial environment, misalignment of this axis w.r.t. the measurement system happens most of the cases, so it can't be ignored.
This tracking reconstruction is computationally more intensive, as it excludes the reported acceleration of the algorithms mentioned before. It will however be highly beneficial in both the reconstruction and analysis phase. A point-tracking technique, presented in part~\ref{ssec:meth:pose}, has been developed to automatically derive the 3D pose of the scanned object from the projection images. Given the simple symmetry no extension to more demanding deep learning techniques~\cite{bui2017x} should be made, however interesting they may be. It is important to note that this technique is not limited to standard circular trajectories. For a high continuous throughput that's aimed at in industrial applications, a conveyor belt trajectory~\cite{Thomas2016line} could be used.
With this prior knowledge the algebraic reconstruction can be improved by including an initial solution that automatically has a correct orientation. Regions can be marked to have a high or low probability of change to improve reconstruction stability~\cite{Marjolein}. These previous developments are not affected by the coordinate base change of the reconstruction described in part~\ref{ssec:meth:cylrec}. The effect of aimed resampling is demonstrated on simulated data in section~\ref{ssec:meth:mtfs} and improved reconstruction algorithms in cylindrical coordinates are applied on a large series of scans in~\ref{ssec:meth:batch}. \\

\section{Methods and materials}

\begin{figure}[ht]
	\centering
	\includegraphics[width=0.5\textwidth]{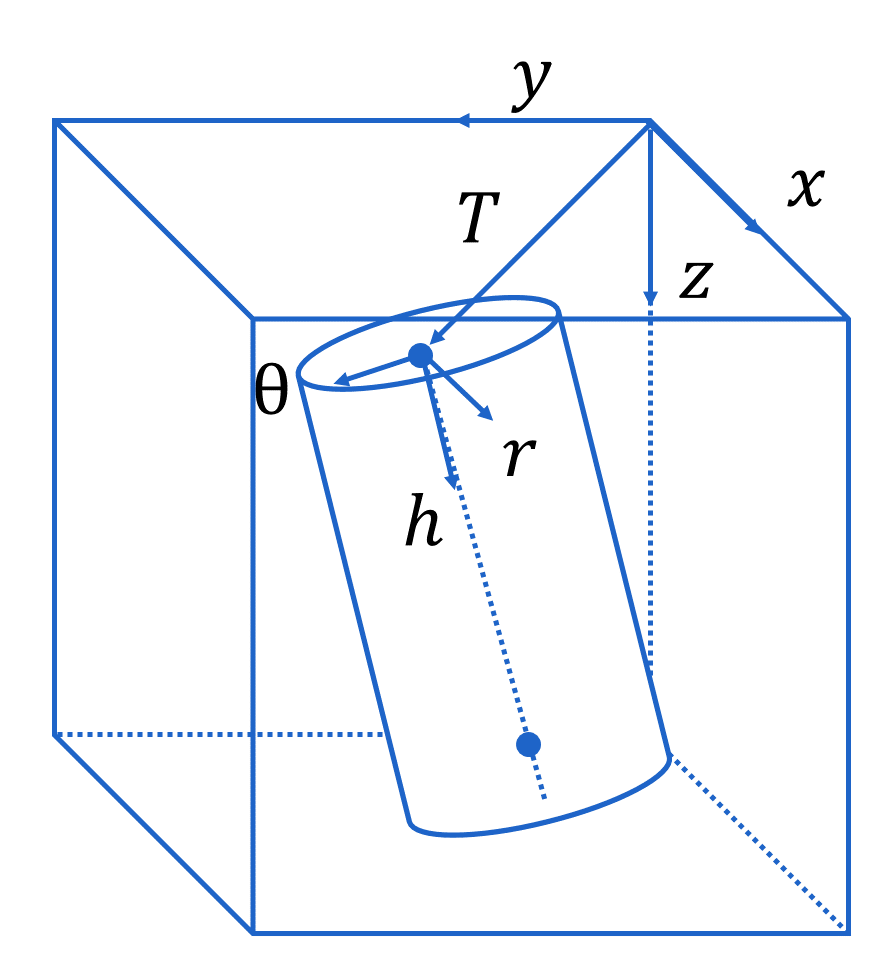}
	\caption{The cylinder can take a free pose in the scanning space.}
	\label{fig:cylinder_tracking}
\end{figure}

Two techniques are used to offer a better framework in the \ac{CT} reconstruction of misaligned cylindrical objects.
\begin{enumerate}
	\item To detect the cylinder orientation from the projection data, a point tracking algorithm is made for general trajectories.
	\item The cylindrical reconstruction is built as an extension of a regular Cartesian \ac{SART}.
\end{enumerate}
The object's pose, illustrated in figure~\ref{fig:cylinder_tracking}, can be captured in six degrees of freedom. Three intrinsic $zxz$ Euler angles $(\alpha, \beta, \gamma)$ are captured in the rotation matrix $R$, while $T$ translates the object along the three spatial coordinates:

\begin{equation}
X' = R X + T
\label{eq:alignment}
\end{equation}

After the alignment from $X = (x,y,z) $ to  $X' = (x', y', z')$ in equation~\eqref{eq:alignment}, a transformation to cylindrical coordinates around the z'-axis is performed on this oriented set of coordinates:

\begin{equation}
\left\{
\begin{aligned}
r &= \sqrt{x'^2 + y'^2} \\
\theta &= \arctan\left(\dfrac{y'}{x'}\right) \\
h &= z' \\ 
\end{aligned}
\right.
\label{eq:cylindrical}
\end{equation}

\subsection{Projection based pose estimation}\label{ssec:meth:pose}

\subsubsection*{Point tracking}

Given a single point in the 3D volume space $(x,y,z)$, the horizontal and vertical projected coordinate $(u,v)$ of the corresponding volume element or \textit{voxel} can be uniquely determined given the view transformations $ \mathcal{V} $ of the source-volume-detector system. The inverse problem of projection-based point tracking is ill-posed as any point lying on the straight line coming from the source point can cast a shadow on the given detector pixel. 
Combining $P$ projections from different directions $ i $ will overcome this problem and leave you with an overdetermined set of non-linear equations due to the conical shape of the beam. 
\begin{equation}
(u, v)_i = \mathcal{V}_i(x,y,z) \qquad i = 1,2, \dots , P
\end{equation}
The Levenberg-Marquardt algorithm~\cite{marquardt1963} yields a solution $ (x,y,z)_S $ that minimizes the sum of squared projection distances. Derivatives of $\mathcal{V}_i(x,y,z)$ can be calculated analytically.
\begin{equation}
(x,y,z)_S = \argmin_{x,y,z}\sum_{i=1}^P\left|(u,v)_i - \mathcal{V}_i(x,y,z) \right|^2 
\label{eq:backtracking}
\end{equation}

\subsubsection*{Axis tracking}

To track the cylinder axis, two 3D points along the axis are sufficient to fix five degrees of freedom, leaving the internal rotation as a last free parameter. In practice these points can be seen by finding the symmetry axis of the cylindrical object on the projections. Due to the symmetric nature of the objects, this algorithm can be executed very fast using tailor-made \ac{GPU} techniques. This technique of projection-based pose estimation may be impossible for objects of lower symmetry.\\

\begin{figure}[ht]
	\centering
	\includegraphics[width=0.9\textwidth]{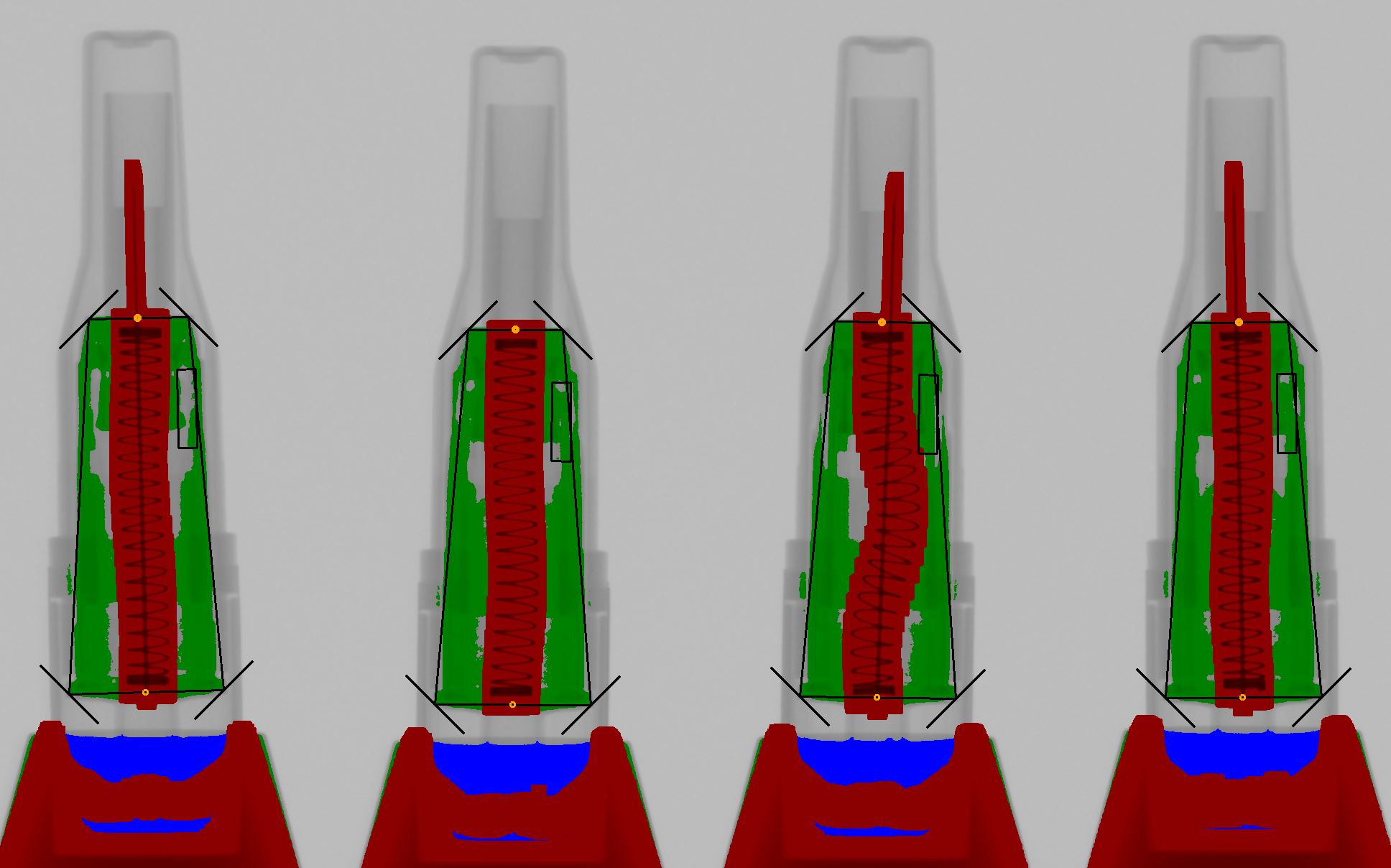}
	\caption{Basic segmentation methods like thresholding and dilation are sufficient to find the projected symmetry axis of the objects.}
	\label{fig:projection_analysis}
\end{figure}

A practical implementation is illustrated in figure~\ref{fig:projection_analysis}. First the most stable symmetric feature is segmented by threshold and dilation techniques. Then the pairwise centres of the four corner pixels are assumed to be on the symmetry axis. These two centres are tracked on each projection and in this way the 3D position of the object can be calculated using equation~\eqref{eq:backtracking}. Two other techniques have been looked into, but although they were not successful in this particular case, they can be of use in other applications. The first one, using the mean and the covariance matrix of the segmented piece, is affected too much by internal variations of the object. The position and angle of the segmented object can also be achieved using the \textit{minAreaRect} function of OpenCV~\cite{itseez2015opencv}. This works by finding the bounding rotated rectangle with minimal area around the segment. Here internal intensity fluctuations at the outer edges affect the outcome as well. 


\subsubsection*{Azimuthal phase}

In some cases it may be necessary to know the azimuthal phase of the object before reconstructing. When a detailed initial solution or weighted volume is used, any internal rotation of the object will produce false features in the reconstruction. 
This last Euler angle $\gamma$ can be derived by monitoring the grey value fluctuations in a region parallel to the projected axis. This will give similar results in function of the projection angle, only phase shifted due to the object's internal rotation and the geometrical angle $\phi_G$ w.r.t. the optical axis of the scanner. 

This perceived phase shift $\phi_P$ can be translated back to the last Euler angle when a correction for the first Euler angles $\alpha$ and $\beta$ is also taken into account. After all, the phase was calculated w.r.t. the original $x$-axis, not the $x'$-axis. 

\begin{equation}
\gamma = \mathrm{\phi_P} - \mathrm{\phi_G} -\arctan\left(\dfrac{\tan(\alpha)}{\cos(\beta)}\right)
\label{eq:azimuthal}
\end{equation}

\subsection{\acs{SART} reconstruction in cylindrical coordinate system}\label{ssec:meth:cylrec}

CT reconstruction seeks to calculate the distribution of attenuation coefficients $\mu$ of an object given the projected X-ray intensities on the detector for a number of projection angles. The basic operation in these methods is back-projection, which tries to invert the set of monochromatic Lambert-Beer equations for all pixel values $p_j$ to a grid of attenuation coefficients $\mu_l$:

\begin{equation}
p_j = -\ln\left[\dfrac{I_j}{I_{0,j}}\right] = \sum_{l=0}^{L} t_{jl}\cdot\mu_l
\label{eq:Beerset-comp}
\end{equation}

The weights $t_{jl}$ represent the intersection length of the ray $j$ passing through the voxel $l$, where $j$ runs over all pixels of all projection images.  
In the \ac{SART} approach, the solution is found by updating each voxel with a correction factor that's found by comparing the actual projection with a simulation on the current state of the volume~\cite{Thomas}. This update step $k$ is performed for each projection in \ac{SART}. The correction factor is accompanied by the relaxation $\lambda$ which dampens the contribution to each update. 

\begin{equation}
\vec{\mu}_m^{(k+1)} = \vec{\mu}_m^{(k)} + \lambda \sum_{j\in\mathrm{OS}}
\dfrac{\left(p_j - \sum_{l=0}^{L}t_{jl} \cdot \mu_l^{(k)} \right)}{\sum_{l=0}^L t_{jl}}
\cdot\dfrac{t_{jm}}{\sum_{j\in\mathrm{OS}} t_{jm}}
\label{eq:SART}
\end{equation}

The projection step is calculated in a pixel driven sampling approach, by tracing the rays between source and detector pixel with an equidistant sampling scheme in Cartesian space. The ray-voxel intersection weights $t_{jl}$ are calculated intrinsically as the number of samplings in each voxel. The current attenuation coefficients are interpolated on the cylindrical grid. This method is also easily extendible to other division schemes. \\


The same strategy can be used for the back-projection step, which is executed in a voxel driven approach. Only the simple coordinate transform ~\eqref{eq:cylindrical} is applied before the corresponding correction term is accessed. The total overhead translates in this case to one cylindrical transformation each time a volume access is performed. Only the alignment matrix needs to be precomputed for equation~\eqref{eq:alignment}. Weighted back-projection, as implemented by Heyndrickx et al.~\cite{Marjolein2016prior}, is adapted similarly by performing the right coordinate transforms. A weight map is prepared in cylindrical coordinates to indicate how severely regions can be subject to change.

\subsection{\acs{MTF} measurements on phantom data}\label{ssec:meth:mtfs}

Due to the way that the reconstruction grid is divided, the number of azimuthal divisions can be changed effortlessly. 
Depending on the application, reducing the number of voxels in a direction that contains broad, smooth structures can be tolerated or may even be necessary to reduce aliasing effects.

\begin{figure}[ht]
	\centering
	\includegraphics[width=\textwidth]{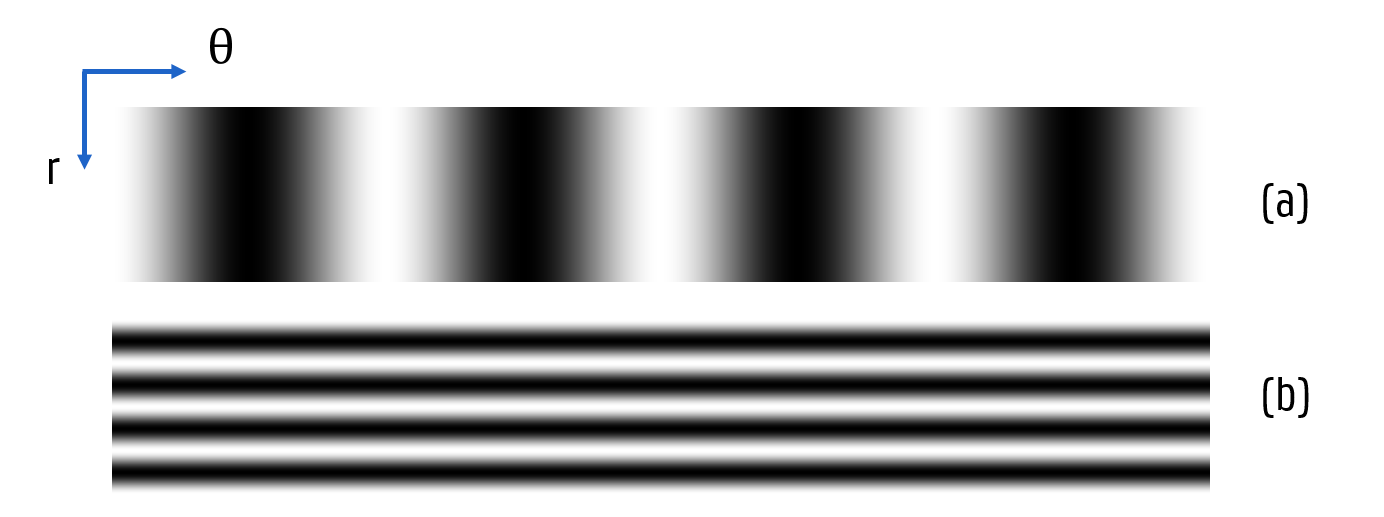}
	\caption[\acs{MTF} Line phantom]{Azimuthal (a) and radial (b) line phantoms, here with $n_l = 4$, are created to test the response of the cylindrical reconstruction method.}
	\label{fig:line_phantom}
\end{figure}

To get a sense of the resolution dependence on the \ac{Nt} and \ac{Nr}, we perform a high detail simulation on samples consisting of radial or azimuthal lines as shown in figure~\ref{fig:line_phantom}. Then the directional \ac{MTF}~\cite{boreman2001modulation} (radial or azimuthal \ac{MTF}) of the reconstruction is determined by measuring the modulation (resp. \acs{MTFR} and \acs{MTFT}) for varying frequency factor, i.e. the number of lines $n_l$ in the phantom of constant size. The modulation $m(n_l)$ for a given frequency $n_l$ is determined by measuring the minimum and maximum reconstructed attenuation coefficients $\mu_{\min} $ and $\mu_{\max} $ in the central slice (to eliminate the effect of cone beam artefacts).

\begin{equation}
m(n_l) = \dfrac{\mu_{\max}(n_l) - \mu_{\min}(n_l)}{\mu_{\max}(n_l) + \mu_{\min}(n_l)}
\label{eq:mtf}
\end{equation}

This value varies between $0$ and $1$, given the positive values of $\mu$. A value of $1$ is expected to be the result of a good response of the reconstruction algorithm to the used phantom frequency. However, one downside to this measure is that it does not punish aliasing effects that alter the line frequency but maintain the modulation depth. A visual inspection is also required to investigate these faulty cases.
In all reconstructions, also the radial dependency of the \ac{MTF} is measured. For the \ac{MTFT} this is simply the \ac{MTF} at each $r$. The radial dependency of the \ac{MTFR} is calculated with a short-range sliding window over the line period. The window width is equal to one period of the phantom. The slide step is equal to the width instead of $1$ voxel, for ease of implementation. \\

The lines are modelled by a cosine function on the respective direction. For the \ac{MTFT}, each slice of the phantom looks like a Siemens star pattern~\cite{gonzalez2015mtf}. The \ac{MTFR} is measured in a phantom of concentric circles.
Highly detailed projections are simulated on these samples of shape $(S_h,S_\theta,S_r) = (32,1608,256)$. 
The frequencies are doubled until a factor of $128~( = S_r / 2)$ is reached. 
The image is projected on 4096 images of dimensions $(W,H) = (2048, 128)$ to reduce the influence of sampling artefacts in the simulation step.  \\

\subsection{Application in batch \acs{CT} scan inspection}\label{ssec:meth:batch}

To illustrate how the developed strategy can be applied in a real production environment, a scan was performed on a series of medical devices with high safety demands.  50 objects have been scanned, reconstructed and analysed for artificially induced internal defects. 
To measure the axis detection precision and to assess the influence of the scanner geometry, the same samples have been scanned repeatedly with different centres of rotation. The pose w.r.t. the rotation stage is constant. The projected horizontal coordinates of the rotation stage are $185.2$, $574.5$, $963.9$, $1353.25$ and $1742.6$ pixels. For the geometry displayed in table~\ref{tab:batch_scan}, the precision is measured as the standard deviation of the pose estimation for the five consecutive scans. This measurement is done over for 50 samples.

\begin{table}[!ht]
	\caption[batch scan parameters.]{Imaging parameters for high speed circular cone beam scans of medical devices}
	\centering
	\begin{tabular}{| l l l c r l |}
		\hline
		&Geometry 		&Circular cone beam 		&				&			&\\
		&				&Source Detector Distance 	& \acs{SDD}		&791		&mm\\
		&				&Source Object Distance 	& \acs{SOD}		&679		&mm\\
		&				&Number of projections		& \Np			&21			&  \\
		&				&Last scan angle			& 				&200$\degree$&	\\
		&Detector		&Varex 1512					&				&			& \\
		&				&Detector size				& (H,W)			&(1536,1944)&\\
		&				&Detector pixel size		& $p_d$			&0.748		&mm\\
		&Sample			&KN-S						& 				&			&\\
		&				&Height						& 				&64			&mm\\
		&				&Radius						& 				&8			&mm\\
		\hline
	\end{tabular}\label{tab:batch_scan}
\end{table}

\section{Results}\label{sec:Results}

\subsection{Characterization on simulated data}\label{ssec:res:mtfs}

\begin{figure}[ht]
	\centering
	\includegraphics[width=\textwidth]{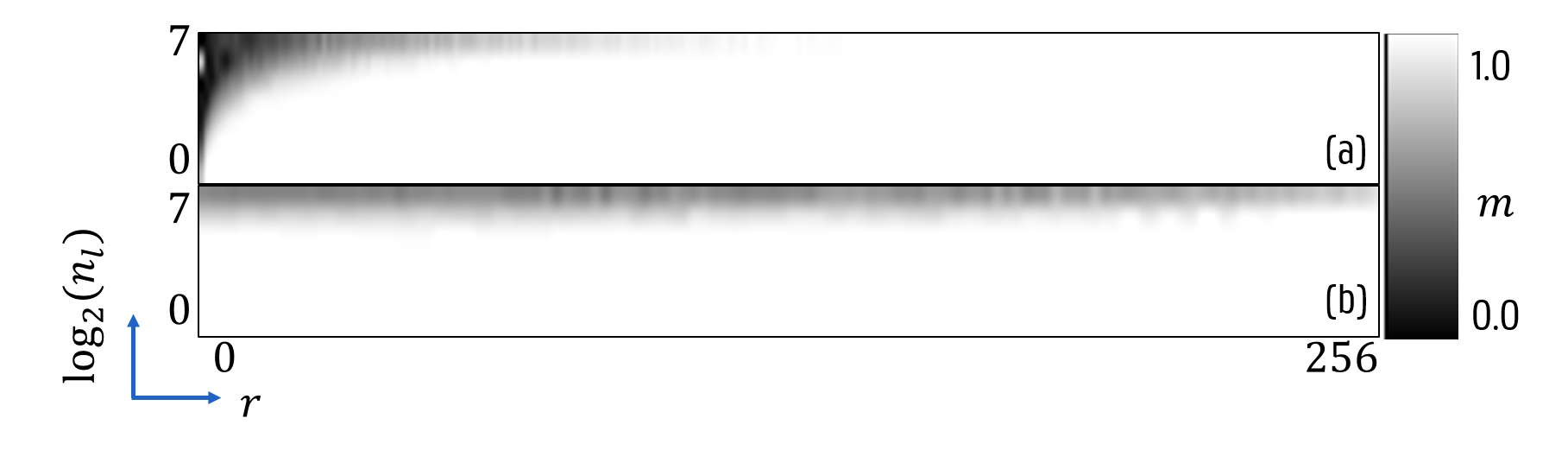}
	\caption[radial dependence \acs{MTF}]{At each radial position $r$ of the reconstruction, the \acs{MTFT} (a) and \acs{MTFR} (b) are measured. Each vertical line represents an \ac{MTF}. The \acs{MTFT} (measured in the Siemens star-like phantom), becomes worse at low radii, not being able to reconstruct the high frequency signals. For the \acs{MTFR}, only a small radial dependence is found.
	}
	\label{fig:mtfs}
\end{figure}

The \ac{MTFT} and \ac{MTFR} at reconstruction shape $(\Nhtr) = (32,512,256)$ are shown in figure~\ref{fig:mtfs}. As expected, the \ac{MTF} becomes worse at high frequent lines. The azimuthal \ac{MTF} shows worse behaviour at small radii. Even though the voxel size is virtually decreased at lower $r$, the actual resolution doesn't automatically follow this trend. 
For the radial \ac{MTF}, only a minor effect of the centre distance can be observed in figure~\ref{fig:mtfs}(b).\\

No variation of the \ac{MTFR} was noted for any \Nt between 1 and 512. Do note that this is a very extreme phantom and the results will probably be a mixed term in real objects.

\begin{figure}[ht]
	\centering
	\includegraphics[width=\textwidth]{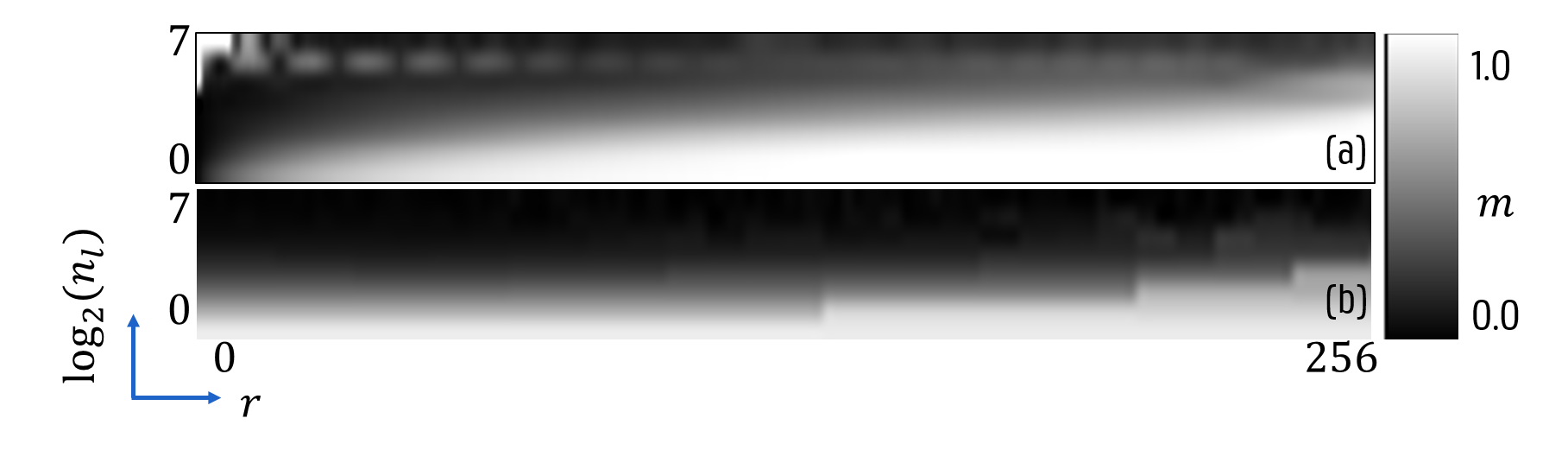}
	\caption[radial dependence \acs{MTF} at fewer projections]{When fewer projections are taken, both \acs{MTFT} (a) and \acs{MTFR} (b) are affected, but they behave similarly to the previous case (figure~\ref{fig:mtfs}). The high \ac{MTFT} at high frequencies and the rise at higher $r$  are most likely caused by reconstruction aliasing effects.}
	\label{fig:mtfs2}
\end{figure}
\begin{figure}[ht]
	\centering
	\includegraphics[width=\textwidth]{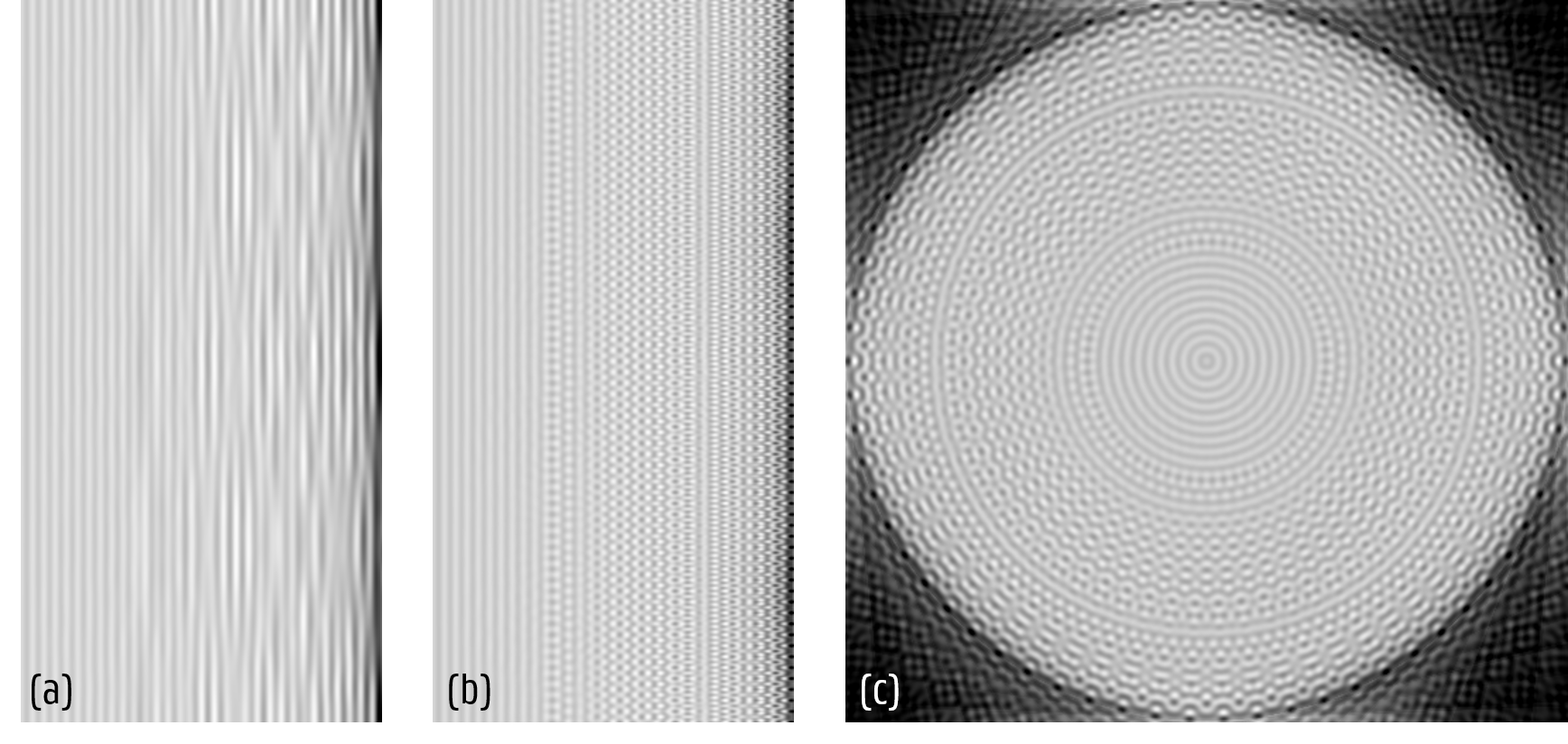}
	\caption[changing \Nt at fewer projections]{Decreasing \Nt in (a) compared to (b) reduces aliasing artefacts when an insufficient amount of projections is given. This can't be done for a Cartesian reconstruction (c), where the results are similar to the reconstruction in (b).}
	\label{fig:mtf_rec}
\end{figure}

The same exercise has been done over for a lower \ac{Np}. Now $67$ instead of $4096$ projections were simulated and the same reconstruction shape was used. Similar results have been noted for $64$ projections, the odd number of $67$ was chosen to avoid any numerical bias. These results are shown in figure~\ref{fig:mtfs2} and the modulation factor is lower than the previous outcome as expected. More variation can be observed for the \ac{MTFR} phantom in the radial direction. When we look at the reconstructed slices in figure~\ref{fig:mtf_rec}, the origin of the seemingly good \ac{MTFR} at high $r$ is revealed. Due to the insufficient amount of projections, aliasing effects appear in the reconstruction. Actually, the quality is better at low $r$ where no azimuthal aliasing shows up at lower frequencies. Decreasing the number of azimuthal divisions \Nt to the order of the projections number gives more truthful results, though the aliasing is still visible. When the same radial resolution is required at a greater distance from the central axis, a higher \Np is required. This requirement remains true for hollow cylinders. Understanding these aliasing effects can help in the development of more advanced sampling strategies. \\

\subsection{Precision of pose estimation}\label{ssec:res:pose}

\begin{figure}[ht]
	\centering
	\includegraphics[width=\textwidth]{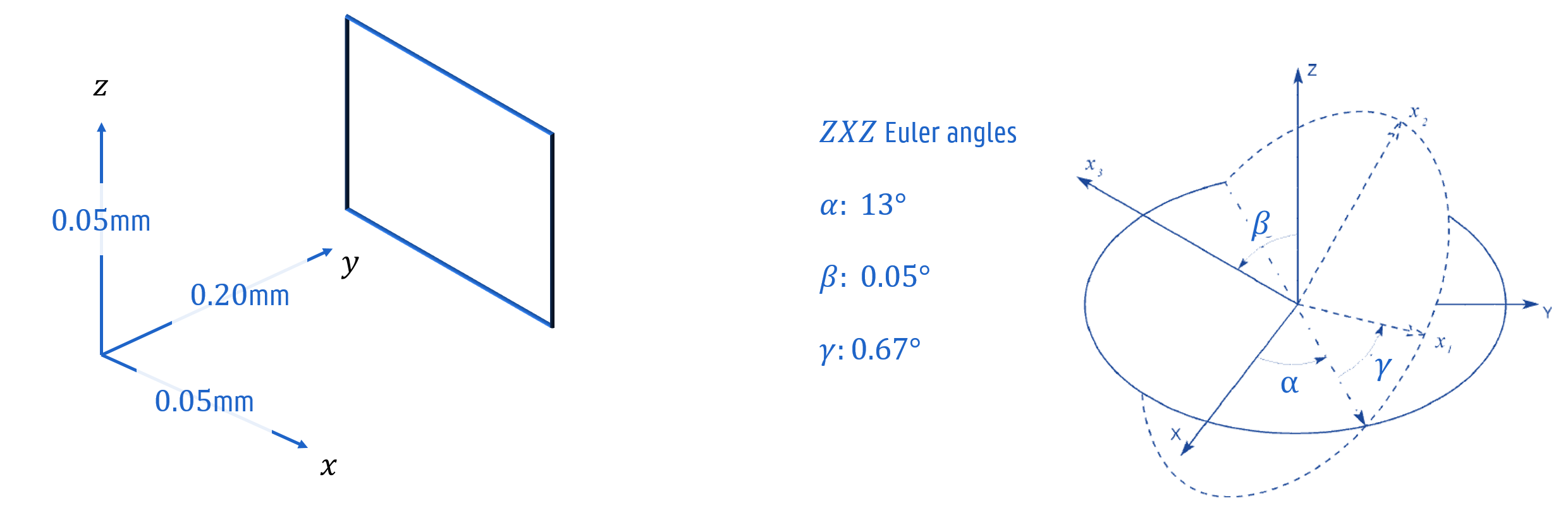}
	\caption[Precision of pose estimation]{The precision of translational and rotational pose estimation. The object's size is given in table~\ref{tab:batch_scan}.}
	\label{fig:pose_precision}
\end{figure}

A scan has been performed in a geometry that resembles an in-line CT inspection set-up. As can be seen in table~\ref{tab:batch_scan}, the objects are placed rather close to the wide detector in order to scan more samples at the same time. For these scans, the axis detection using the extremal corner pixels took only $0.23$s.
The \textit{minAreaRect} technique took $8$s on average, so this alternative was not used. 
The precision of the pose estimation, derived as explained in paragraph~\ref{ssec:meth:batch}, is shown in figure~\ref{fig:pose_precision}. The translational precision is better in the transversal directions. Due to the small $\beta$ angle, $\alpha$ seems to suffer bad precision, though this is only a mathematical issue due to the used Euler angle convention. The difference in $x$ and $y$ might be attributed to the limited angle of $200^\circ$ in which projections were taken. The overall good precision is necessary to assure a qualitative reconstruction. \\

The practical difficulties of this method are present in the segmentation phase, as demonstrated in figure~\ref{fig:projection_analysis}. The projected intensity won't be constant for a given cylindrical object. The total attenuation will be less at the cylinder edges, where a smaller amount of material was traversed. A simple threshold technique may be affected by disturbing features in the object. In this case the sample holder at the image bottom and the metal components reduce the projected intensity. To remove this degeneracy in grey values, the other features are separately thresholded and dilated before removing them from the collection of foreground pixels. The sample holders assure already a rough estimate of the positions so some regions can already be excluded.

\subsection{Use of initial volume and back-projection weights in cylindrical reconstruction algorithms}\label{ssec:res:adv}

\begin{figure*}[ht]
	\centering
	\subfigure[]{\includegraphics[width=0.22\textwidth]{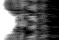}
		\label{fig:batch_simple_rec}}
	\subfigure[]{\includegraphics[width=0.22\textwidth]{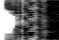}
		\label{fig:batch_init_rec}}
	\subfigure[]{\includegraphics[width=0.22\textwidth]{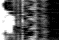}
		\label{fig:batch_init_weight_rec}}
	\subfigure[]{\includegraphics[width=0.22\textwidth]{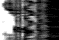}
		\label{fig:batch_weight_rec}}
	\caption[advanced batch reconstructions]{A cropped $r \theta$ slice of four different cylindrical reconstructions, severely influenced by a bright metal spring and needle. In~\subref{fig:batch_simple_rec} the \ac{SART} reconstruction is executed without prior knowledge. \subref{fig:batch_init_rec} and ~\subref{fig:batch_init_weight_rec} use an initial volume, while weighted back-projection is applied in~\subref{fig:batch_init_weight_rec} and~\subref{fig:batch_weight_rec}.}
	\label{fig:batch_recs}
\end{figure*}

The added value of the initial solution and the weighted back-projection are investigated for the batch scan with settings listed in table~\ref{tab:batch_scan}. The number of voxels in the cylindrical reconstruction is $(\Nhtr) = (240,40,60)$ for a virtual voxel size of $0.312$mm.  Reconstruction times changed from $0.30$s for a normal one, to $0.52$s for a weighted back-projection reconstruction.
In figure~\ref{fig:batch_recs}, the results are illustrated for each distinct reconstruction strategy:
\begin{enumerate}[(a)]
	\item The object is reconstructed without aid of initial volume or weighted back-projection.
	\item Include an initial solution, which is considered to be the dangerous approach in the case of missing components. The iterative algorithm might not remove these sufficiently.
	\item Initial solution combined with weighted back-projection. Some predefined regions are given higher weights to change their attenuation value.
	\item Only weighted back-projection: the algorithm starts from a completely empty volume.
\end{enumerate}

From figure~\ref{fig:batch_recs}, one can derive a number of preliminary conclusions. When no prior info is included, the slices have low contrast and sharpness. This will negatively affect the later analysis phase where image characteristics are converted to binary results (present / missing). In sub-figure~\ref{fig:batch_init_rec} a higher sharpness is perceived compared to~\ref{fig:batch_simple_rec}, but this also illustrates an important aspect of the algorithm: when the volume is badly aligned with the initial solution, some ghost effects may appear (particularly in the azimuthal direction). The weighted approaches in figures~\ref{fig:batch_init_weight_rec} and~\subref{fig:batch_weight_rec} improve the contrast, while the latter doesn't include an initial solution. \\

\begin{figure}[ht]
	\centering
	\includegraphics[width=\textwidth]{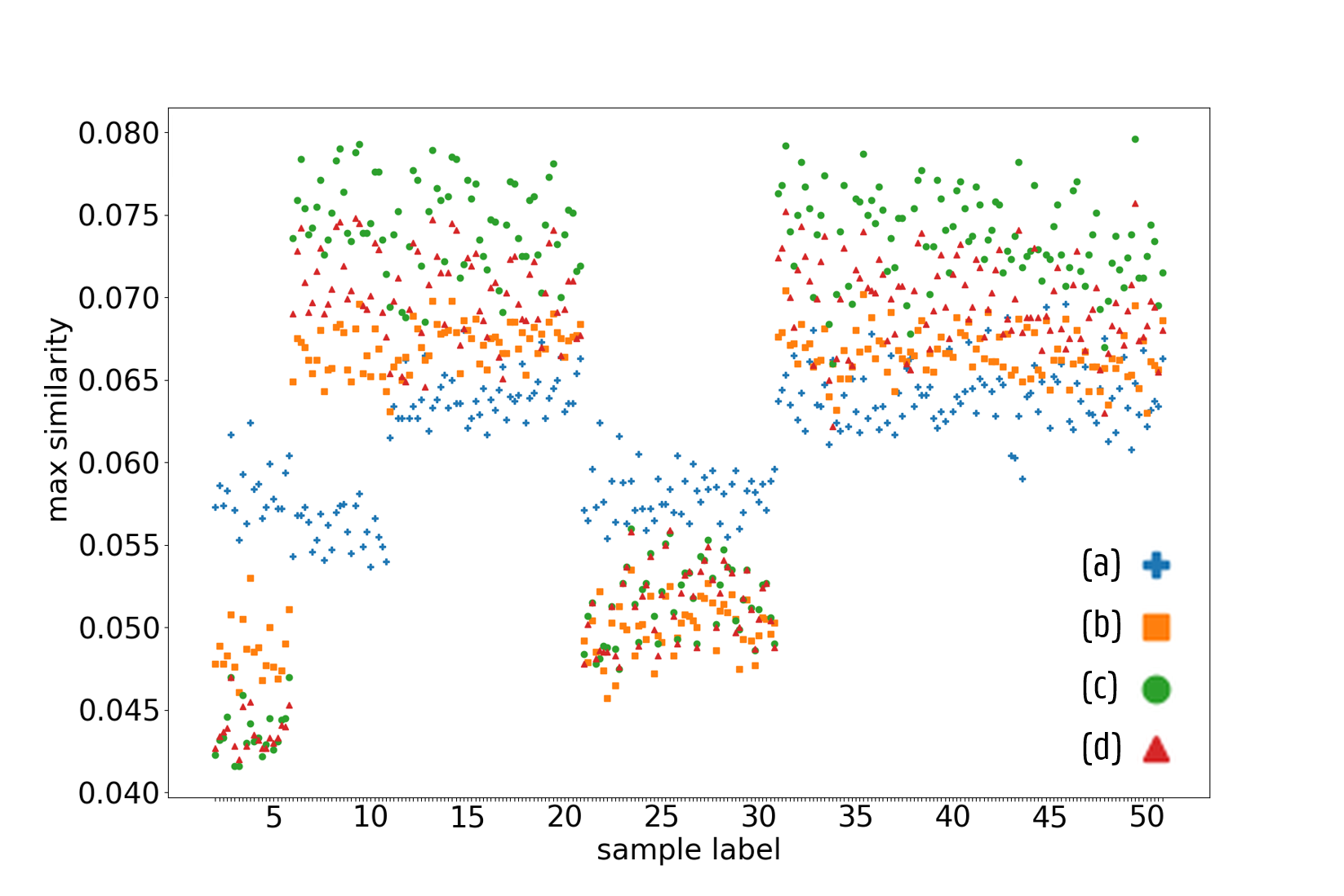}
	\caption[Improved contrast reconstruction]{When the reconstruction contrast is improved, it becomes easier to spot a missing component. The labels (a-d) correspond to the ones given in figure~\ref{fig:batch_recs}.}
	\label{fig:batch_improved_contrast}
\end{figure}

The results can also be observed in the test measuring the piece presence, described earlier. Figure~\ref{fig:batch_improved_contrast} shows the distribution of this measure for the different methods. It is clear that the difference between the groups has increased most when an initial volume and weights are included. This clear distinction is necessary to increase the detection rate of the internal defects without discarding safe devices. 

\section{Conclusion}

In this work two techniques have been developed in the scope of high-throughput \ac{CT} inspection of cylindrical objects.
These are necessary to reduce scan time while maintaining reasonable image quality. A pose estimation technique for cylindrical objects has been implemented that uses only basic segmentation algorithms on the projected images. The axis extraction using the corner information was found to be both fast and precise. Reconstruction was performed directly in the aligned cylindrical coordinate system.

The \ac{MTF} characterization shows how reliably the scanner can reconstruct different frequencies of the (phantom) object. The artificially high number of azimuthal divisions at low radius should only be interpreted as a mathematical construct. The actual resolution corresponds more to the Cartesian voxel size. 
The number of divisions could be matched better with the actual content of the object. When the components only have a low frequency in the azimuthal direction, the user may choose to drop the number of divisions, or similarly for other directions. This reduces the reconstruction time.

The aligned reconstruction allows the inclusion of more advanced techniques when a series of same-type objects is scanned. An initial volume can be used to achieve faster convergence. Weighted reconstruction adds stability to the algorithm and a higher sensitivity to changes.

The methods work for scans of cylindrical objects and can be easily added to existing frameworks. The \ac{SART} reconstruction technique is not mandatory, any iterative method can benefit from alignment. The coordinate change yields equal reconstruction quality in the most simple case. The adaptation to the symmetry of the objects allows that more advanced techniques are added to improve image quality.

\section*{Acknowledgements}
This research is funded by the imec ICON project iXCon (Agentschap Innoveren en Ondernemen project nr. HBC.2016.0164). The work by M. Heyndrickx was funded by the Research Foundation Flanders (FWO) [grant number 1S 215 16N].

The authors would like to thank the manufacturer for providing the samples with introduced faults and also
TESCAN XRE for acquiring part of the projection data (\url{www.tescan.com}).

\bibliographystyle{elsarticle-num} 
\bibliography{citations}   

\end{document}